\documentclass{iau}
\usepackage{graphicx}

\title[Constraining FeLoBAL Outflows] %% give here short title %%
{FeLoBAL Outflow Variability Constraints from Multi-Year Observations}

\author[McGraw et al.]   %% give here short author list %%
{Sean M. McGraw$^1$,
Joseph C. Shields$^1$,
Frederick W. Hamann$^2$,
Daniel M. Capellupo$^3$,
Sarah C. Gallagher$^4$,
 \and William N. Brandt$^5$}

\affiliation{$^1$Department of Physics and Astronomy, Ohio University, Athens, Ohio, United States \\[\affilskip]
$^2$University of Florida, Gainesville, FL, United States \\[\affilskip]
$^3$ Tel Aviv University, Tel Aviv, Israel \\[\affilskip]
$^4$ University of Western Ontario, London, ON, Canada \\[\affilskip]
$^5$ Pennsylvania State University, University Park, PA, United States}

\pubyear{2014}
\volume{304}  %% insert here IAU Symposium No.
\pagerange{119--126}
\jname{Multiwavelength AGN Surveys and Studies}
\editors{A.C. Editor, B.D. Editor \& C.E. Editor, eds.}
\begin{document}

\maketitle

\begin{abstract} The physical properties and dynamical behavior of Broad Absorption Line (BAL) outflows are crucial themes in understanding the connections between galactic centers and their hosts.  FeLoBALs (identified with the presence of low-ionization Fe II BALs) are a peculiar class of quasar outflows that constitute $\sim1\%$ of the BAL population.  With their large column densities and apparent outflow kinetic luminosities, FeLoBALs appear to be exceptionally powerful and are strong candidates for feedback in galaxy evolution.  We conducted variability studies of 12 FeLoBAL quasars with emission redshifts $0.69 \leq z \leq 1.93$, spanning both weekly and multi-year timescales in the quasar's rest frame.  We detected absorption-line variability from low-ionization species (Fe II, Mg II) in four of our objects, with which we established a representative upper limit for the distance of the absorber from the supermassive black hole (SMBH) to be $\sim$20 parsecs.  Our goals are to understand the mechanisms producing the variability (e.g. ionization changes or gas traversing our line of sight) and place new constraints on the locations, structure, and kinetic energies of the outflows. 

\end{abstract}

\firstsection % if your document starts with a section,
\section{Introduction}

Quasar outflows are believed to be a ubiquitous phenomenon within luminous AGN in distant galaxies, however the detailed physics of the outflows is not well understood.  Conducting variability studies of Broad Absorption Lines (BALs), the signatures of outflowing gas, in quasars can prove useful in placing constraints on the physical properties of the gas. 

The motivation behind understanding the physics of outflows is fueled by the overall picture of galaxy evolution and the connection between the central SMBH and its surrounding host.  It is clear that the SMBH mass correlates with host galaxy bulge properties (e.g., the $M-\sigma$ relation), however what is not apparent is to what extent quasar outflows contribute to this connection.  There is evidence of quasar outflows having a direct impact on star formation in the host galaxy (\cite[Farrah et al. 2012]{Farrah_etal12}), yet to probe this feedback mechanism further will require firm observational constraints on the outflow's properties.

Iron Low Ionization BALs, or FeLoBALs, form a rare subclass of BAL quasars ($\sim$ 1\%) that distinguish themselves by showing BALs in Fe II and Fe III in addition to typical low and high ionization species found in the rest of BAL quasars.  Theoretical models on the origin of FeLoBAL outflows are quite varied, as are their implications for the AGN-host galaxy connection.  \cite[Faucher-Gigu\`ere et al. (2012)]{Faucher_etal12} suggest that a quasar blast wave interacts with the ISM at $\sim$ 1 kpc from the SMBH, creating cloudlets that form the absorption lines.  \cite[Murray et al. (1995)]{Murray_etal95}, on the other hand, model a radiation-pressure-driven BAL wind that arises from the accretion disk on sub-parsec scales from the SMBH, a situation that might explain  all the BAL quasar subclasses.

Variability studies of FeLoBAL quasars are rare.  \cite[Hall et al. (2011)]{Hall_etal11} observed absorption line variability in 1 FeLoBAL, constraining the absorber's distance to be between 1.7 and 14 pc from the SMBH.  \cite[Vivek et al. (2012)]{Vivek_etal12} investigated 5 FeLoBALs, detecting changes in absorption from Al III and Fe III in one of their objects.  More constraints need to be placed on FeLoBAL outflows in order to have a clearer understanding of their origins, structure, and impact on the surrounding host galaxy.

\section{Results}  

We have obtained preliminary results from a variability study of 12 FeLoBAL quasars (0.69 $\leq z \leq$ 1.93) observed over timescales ranging from $\sim$10.3 days to $\sim$7.03 years in the quasar's rest frame.  The wavelength coverage across our sample spanned between $\sim$3800 -- 7250 \AA, with up to 5 epochs available for a given object.  The sample included spectra obtained using the MDM 2.4m Hiltner telescope at Kitt Peak, AZ, along with data from the 9.2m HET telescope at McDonald Observatory and SDSS Data Release 7.

Our initial analysis shows absorption-line variability mainly from Fe II and Mg II in 4/12 FeLoBALs over restframe timescales of $\sim$0.5 to $\sim$6.4 years.  Regions of variability occur both in portions of lines and across entire BAL troughs.  In one of our objects we have detected correlated absorption-line variability in Fe II from both ground (UV 1,2,3) and excited (UV 62,63,78) state transitions, an observation that is uncommon as most variability studies detect changes from resonance lines only.  Variability was quantified in each object by comparing flux differences between two spectra to the propagated noise of the resulting difference.  We defined a preliminary criterion for variability to be that the flux difference between two epochs rises $\geq$ 3$\sigma$ above the noise across $\geq$ 3 consecutive pixels in order to be significant.

In order to place new physical constraints on FeLoBAL outflows, the cause of the variability becomes important.  Two scenarios for the origin of variability in quasar outflows are a change in ionization within the gas or a change of covering fraction due to transverse motions across the line of sight.  Using a typical crossing timescale of $\Delta t\sim3.08$ years for our sample, we can constrain the absorber to be $\lesssim$ 20 pc from the SMBH with crossing speeds $\gtrsim$ 500 km/s.  Categorizing the variability in our sample becomes challenging due to the large number of Fe II multiplets blended together and the possibility that multiple transitions from Fe II and/or other ions contribute to any given absorption line.  A forthcoming paper will discuss individual objects in our sample along with the implications of the variability for constraining FeLoBAL outflows.

\end{document}